\documentclass[pre,twocolumn,showpacs,floatfix,amsmath,amssymb]{revtex4-1}

\usepackage{graphicx, color}
\usepackage{amsmath,amssymb}
\usepackage{verbatim}

\newcommand{\tableone}{
	\begin{table*}[t]
		\begin{ruledtabular}
			\begin{tabular}{r l l l l l l l l }
				\multicolumn{1}{c}{} & \multicolumn{2}{c}{$G=1$} & \multicolumn{6}{c}{$G=3$} \\
				\cline{2-3}
				\cline{4-9}
				$N$ & $\omega^{*} = \frac{\pi}{8N}$ & $\omega^{*}$ (OBS) & $\omega^{*} =
				\frac{5\pi}{24N}$ & $\omega^{*}$ (OBS) & $\omega_{E} = \frac{11\pi}{12N}$ &
				$\omega_{E}$ (OBS) & $\omega_{2} = \frac{3\pi}{2N}$ & $\omega_{2}$ (OBS) \\
				\hline \hline
				32  & $1.2 \times 10^{-2}$ & $1.4(1) \times 10^{-2}$ & $2.0 \times
				10^{-2}$ & $1.8(6) \times 10^{-2}$ & $9.0 \times 10^{-2}$  & $9.2(6)
				\times 10^{-2}$  & $1.5 \times 10^{-1}$  & $1.3(1) \times 10^{-1}$ \cr
				64  & $6.1 \times 10^{-3}$ & $6.2(5) \times 10^{-3}$ & $1.0 \times
				10^{-2}$ & $8.4(4) \times 10^{-3}$ & $4.5 \times 10^{-2}$  & $4.6(2)
				\times 10^{-2}$  & $7.4 \times 10^{-2}$  & $6.5(3) \times 10^{-2}$ \cr
				128 & $3.0 \times 10^{-3}$ & $2.8(1) \times 10^{-3}$ & $5.1 \times
				10^{-3}$ & $4.2(1) \times 10^{-3}$ & $2.25 \times 10^{-2}$ &
				$2.27(3)\times 10^{-2}$  & $3.6 \times 10^{-2}$  & $3.2(2) \times 10^{-2}$ \cr
				256 & $1.5 \times 10^{-3}$ & $1.4(1) \times 10^{-3}$ & $2.5 \times
				10^{-3}$ & $1.9(1) \times 10^{-3}$ & $1.12\times 10^{-2}$  & $1.12(1)
				\times 10^{-2}$ & $1.8 \times 10^{-2}$ & $1.63(2) \times 10^{-2}$ \cr
				512 & $7.7 \times 10^{-4}$ & $6.9(3) \times 10^{-4}$ & $1.28 \times
				10^{-3}$ & $9.5(3) \times 10^{-4}$ & $5.6 \times 10^{-3}$ & $5.65(4)
				\times 10^{-3}$ & $9.2 \times 10^{-3}$ & $8.1(1) \times 10^{-3}$ \cr
			\end{tabular}
		\end{ruledtabular}

		\caption{\label{table:1} Calculated and observed (OBS) frequencies for various lengths
			of the DNA for force amplitudes $G=1$ and $G=3$. The digits in bracket represent
			the uncertainity of the last decimal place. 
		}
	\end{table*}
}

\newcommand{\figone}{
\begin{figure}[t]
	\centering
	\includegraphics[width=3.0in]{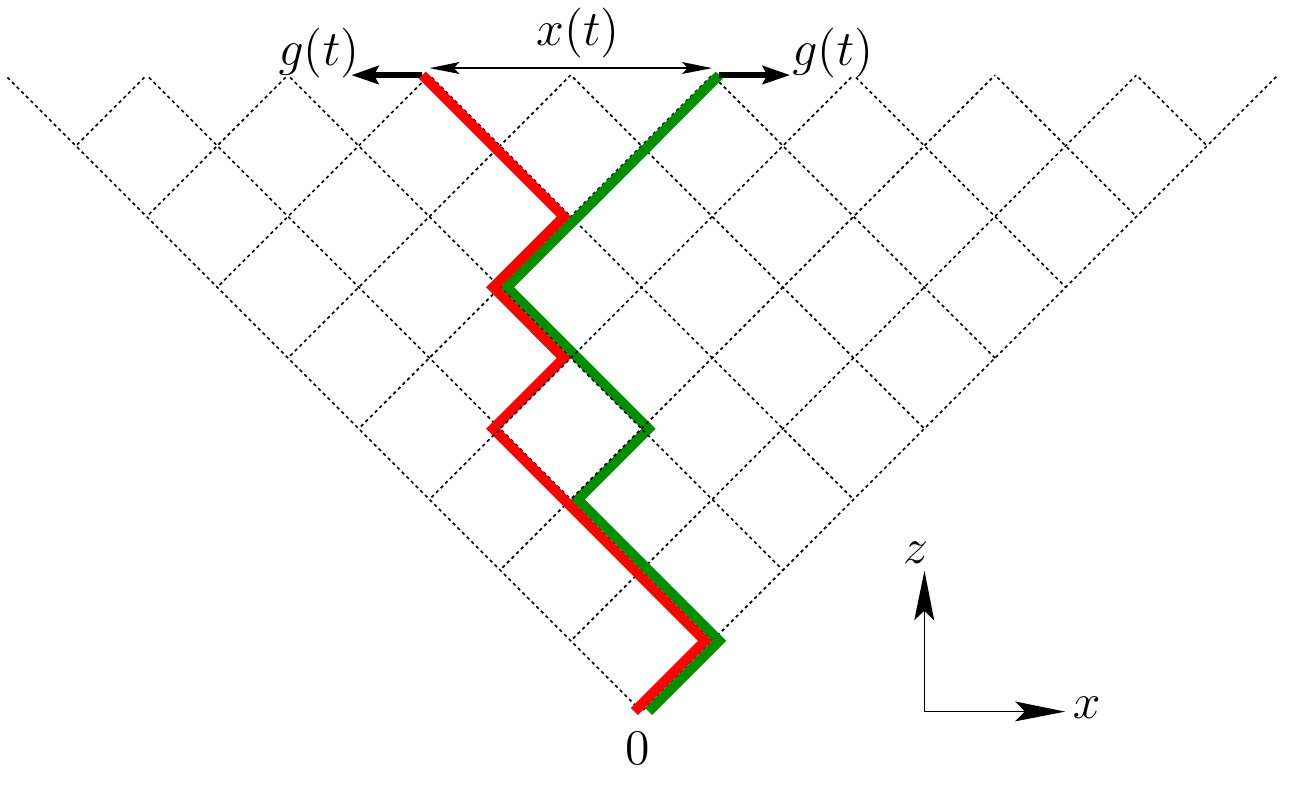}

	\caption{ \label{fig:1} (Color online) Schematic diagram of the model. The strands of the DNA
		are shown by thick solid lines. The end monomers of the strands are pulled along the $x$
		direction with a periodic force $g(t) = G \left| \sin (\omega t) \right|$. The separation
		between the end monomers, $x(t)$, follows the external force $g(t)$ with a lag.
	}

\end{figure}
}

\newcommand{\figtwo}{
\begin{figure}[t]
	\centering
	\includegraphics[width=3.4in]{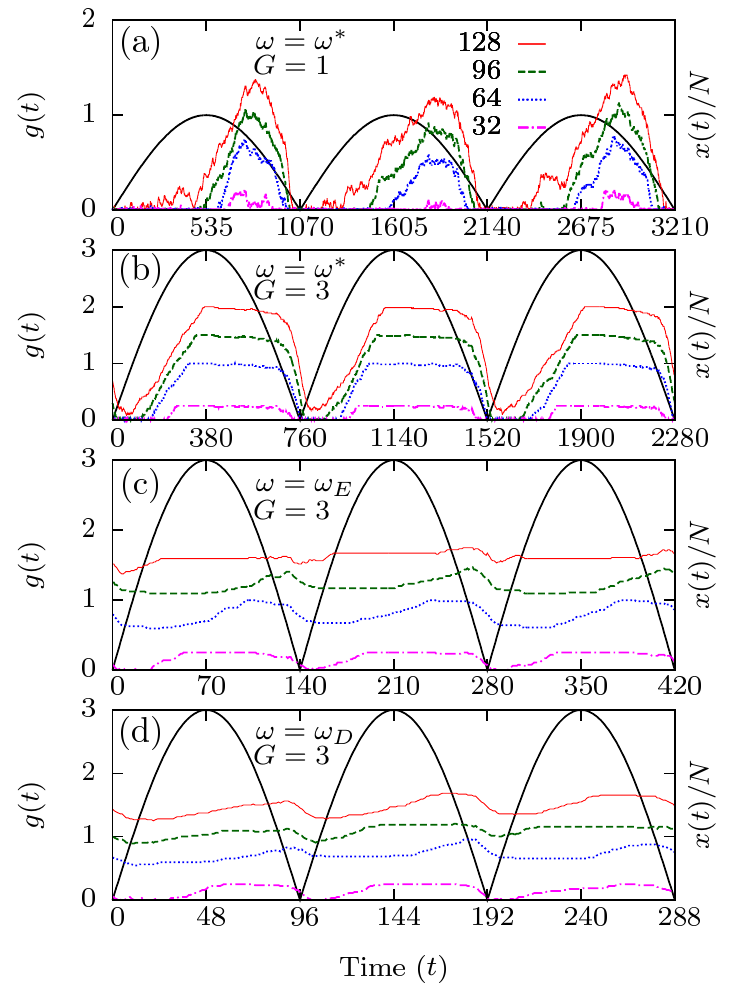}

	\caption{ \label{fig:2} (Color online) (a) The time variation of the external force $g(t)$ for
		frequency $\omega = \omega^{*} = 2.94 \times 10^{-3}$ at force amplitude $G=1$. Various
		lines represent the scaled extension, $x(t)/N$, of different monomers for the DNA of length
		$N=128$. Parts (b), (c), and (d) are the same as (a) for $G=3$ for frequencies $\omega =
		\omega^{*} = 4.2 \times 10^{-3}$, $\omega = \omega_{E} = 2.24 \times 10^{-2}$, and $\omega
		= \omega_{D} = 3.27 \times 10^{-2}$. Here $\omega^{*}(G)$ is the frequency at which
		$A_{loop}$ is maximum, and $\omega_D$ and $\omega_E$ represent the frequencies marked by
		points $D$ and $E$, respectively, in Fig.~\ref{fig:3}(a).
	}

\end{figure}
}

\newcommand{\figfour}{
\begin{figure}[t]
	\centering
	\includegraphics[width=3.5in]{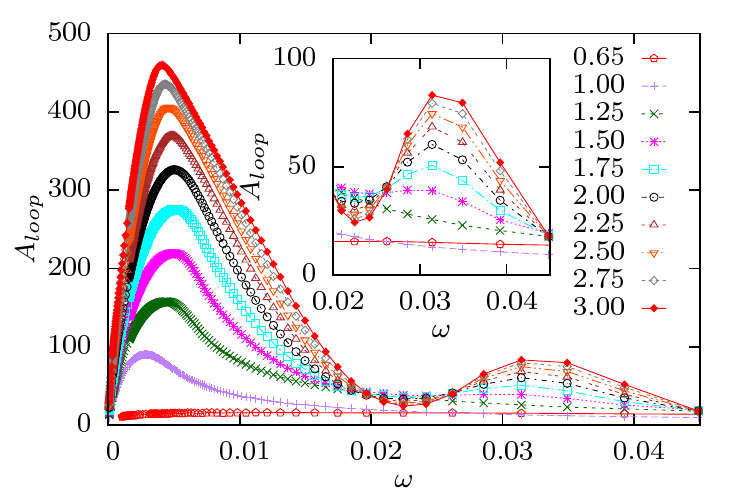}

	\caption{ \label{fig:4}  (Color online) Area of the hysteresis loop, $A_{loop}$, as a
		function of frequency $\omega$ of the periodic force for various force amplitudes $G$
		for the DNA of length $N=128$. The loop area for $G=0.65$ is scaled by a factor of $10$
		to make it visible in the scale. The inset shows the second peak that appears at high
		frequencies.
	}

\end{figure}
}

\newcommand{\figthree}{
\begin{figure}[t]
	\centering
	\includegraphics[width=3.5in]{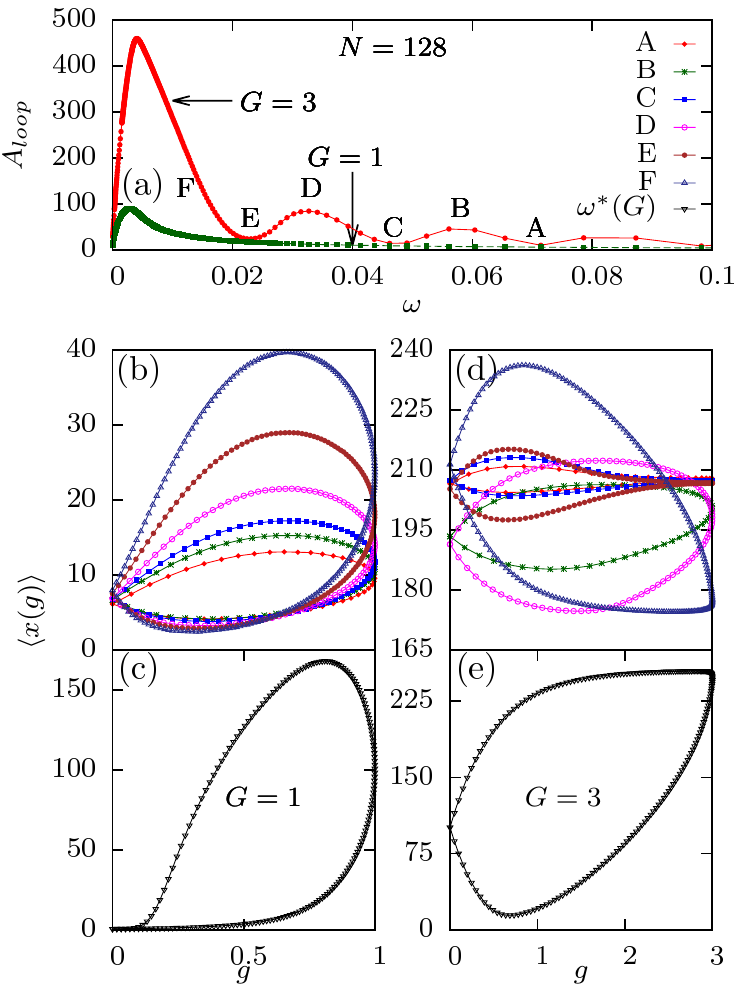}
		
	\caption{\label{fig:3} (Color online) (a) Area of the hysteresis loop $A_{loop}$ as a
		function of $\omega$ in the high frequency range for the force amplitude $G=1$ (filled
		squares) and $G=3$ (filled circles). The length of the DNA is $N=128$. (b) The force
		$g$ versus extension $\langle x(g) \rangle$ curves averaged over $10^4$ cycles for
		$G=1$ at frequencies $\omega_{A} = 7.14 \times 10^-2$, $\omega_{B} = 5.61 \times
		10^{-2}$, $\omega_{C} = 4.62 \times 10^{-2}$, $\omega_{D} = 3.27 \times 10^{-2}$,
		$\omega_{E} = 2.24 \times 10^{-2}$, and $\omega_{F} = 1.57\times 10^{-2}$, indicated,
		respectively, by points $A-F$ in (a). (d) Same as (b) for $G=3$.  Plots (c) and (e)
		shows the hysteresis curves having the maximum loop area at frequencies
		$\omega^{*}(G=1) = 2.8 \times 10^{-3}$ and $\omega^{*}(G=3) = 4.2 \times 10^{-3}$
		respectively.
	}

\end{figure}
}

\newcommand{\figfive}{
\begin{figure*}[t]
	\centering
	\includegraphics[width=6.0in]{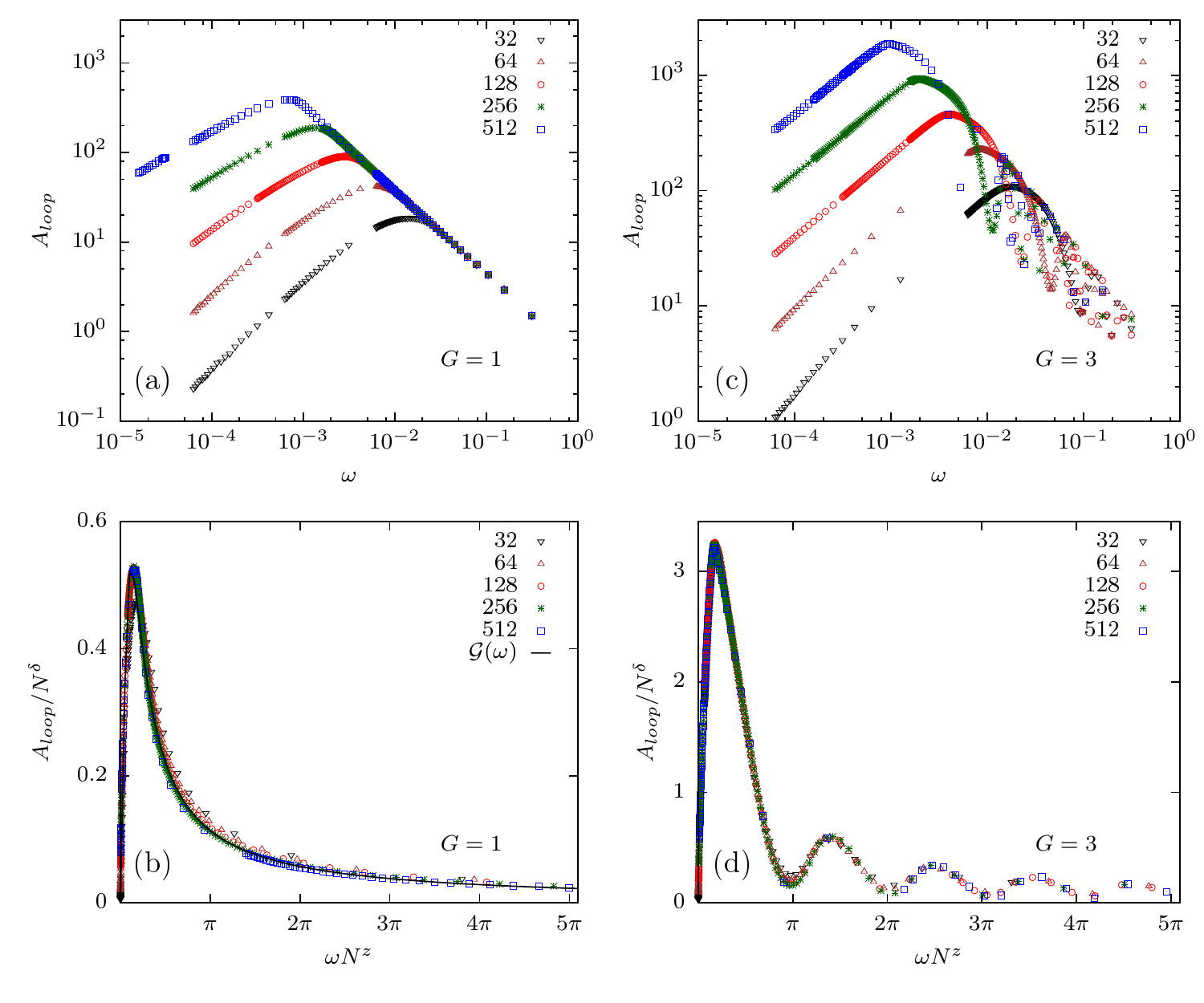}

	\caption{ \label{fig:5} (Color online) Area of the hysteresis loop, $A_{loop}$, for the DNA
		of lengths $N=32,64,128,256,512$, as a function of frequency $\omega$ for force
		amplitudes (a) $G=1$, and (c) $G=3$ in log-log scale. Plots (b) and (d) represent the
		collapse of data shown in (a) and (c), respectively. The values of the exponents are
		close to $\delta=1$ and $z=1$. For $G=1$, the scaling curve $\mathcal{G}(\omega) = A
		G^{\alpha}\omega^{\beta} / (\omega^{1+\beta} + B^2)$ with fitting parameters $A=0.36$ and
		$B=0.30$ is also plotted in (b).  }

\end{figure*}
}

\newcommand{\figsix}{
\begin{figure}[t]
	\centering
	\includegraphics[width=3.4in]{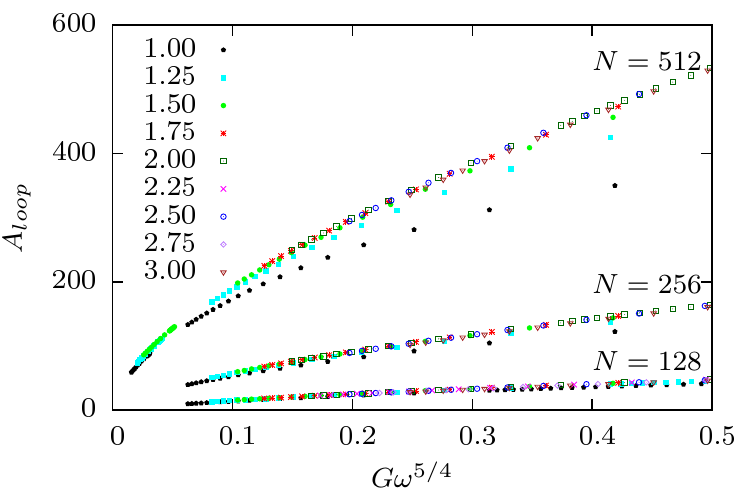}
		
	\caption{\label{fig:6} (Color online) Area of the hysteresis loop $A_{loop}$  versus $G
		\omega^{5/4}$ for the DNA of lengths $N=128$, 256, and 512 at low frequencies for various
		$G$ values. 
	}

\end{figure}
}

\begin{document}

\title{Unzipping DNA by a periodic force: Hysteresis loop area and its scaling}

\author{Rajeev Kapri}
\email{rkapri@iisermohali.ac.in}
\affiliation{Department of Physical Sciences, Indian Institute of Science Education and
Research Mohali, Sector 81, Knowledge City, S. A. S. Nagar, Manauli PO 140306, India.}

\begin{abstract}

	Using Monte Carlo simulations, we study the hysteresis in the unzipping of double stranded
	DNA whose ends are subjected to a time dependent periodic force with frequency ($\omega$)
	and amplitude ($G$). For the static force, i.e., $\omega \to 0$, the DNA is in equilibrium
	with no hysteresis.  On increasing $\omega$, the area of the hysteresis loop initially
	increases and becomes maximum at frequency $\omega^{*}(G)$, which depends on the force
	amplitude $G$.  If the frequency is increased further, we find that for lower amplitudes
	the loop area decreases monotonically to zero, but for higher amplitudes it has an
	oscillatory component.  The height of subsequent peaks decreases and finally the loop area
	becomes zero at very high frequencies. The number of peaks depends on the length of the
	DNA.  We give a simple analysis to estimate the frequencies at which maxima and minima
	occur in the loop area.  We find that the area of the hysteresis loop scales as $1/\omega$
	in the high frequency regime whereas, it scales as $G^{\alpha} \omega^{\beta}$ with
	exponents $\alpha =1$ and $\beta = 5/4$ at low frequencies.  The values of the exponents
	$\alpha$ and $\beta$ are different from the exponents reported earlier based on the
	hysteresis of small hairpins. 

\end{abstract}


\date{\today}

\maketitle

\section {Introduction}

The advent of single molecule manipulation techniques has opened up a new vista in the field of
biophysics. Using these techniques, it is now possible to exert mechanical force, in the
pico-newton range, on an individual molecule giving important information about molecular
interactions~\cite{Ritort2006}. If a mechanical force is exerted on the strands of double
stranded DNA (dsDNA), it unzips when the force exceeds a critical
value~\cite{Bhattacharjee2000,Lubensky2000,Sebastian2000,Bockelmann2002,Danilowicz2004}.  Below
this critical value, the DNA is in the zipped phase, while above it, the DNA is in the unzipped
phase.  Unzipping of a dsDNA has biological relevance. It is an initial step in processes such
as DNA replication and RNA transcription, where the external force is exerted by
enzymes~\cite{Watson2003}. The unzipping transition has been studied, both theoretically and
experimentally, over 15 years and many important results have been established (see
~\cite{Bhattacharjee2000,Lubensky2000,Marenduzzo2001,Marenduzzo2002,Bockelmann2002,Danilowicz2004,Kapri2004,Kapri2006,Kapri2007,Kapri2008,Kapri2009,Kumar2010}
and references therein). In recent years, the focus has been shifted to study the hysteresis in
unbinding and rebinding of biomolecules~\cite{Hatch2007,Friddle2008,Tshiprut2009} because it
can provide useful information on the kinetics of conformational transformations, the potential
energy landscape, and it can be used in controlling the folding pathway of a single molecule
\cite{Li2007}. 

More recently, the behavior of DNA under a periodic force has been explored by using Langevin
dynamics simulation of an off-lattice coarse grained model for a short DNA of $N=16$ base pairs
\cite{Mishra2013,Kumar2013}. It was found that there exists a dynamical phase transition in
which the DNA can be taken from the zipped state to an unzipped state via a new dynamic state.
It was shown that at low frequencies, the area of the hysteresis loop, $A_{loop}$, which
represents the energy dissipated in the system scales with $G^{\alpha} \omega^{\beta}$, where
$G$ is the amplitude, and $\omega$ is the frequency of the oscillating force. The scaling
exponents $\alpha = \beta = 1/2$, were found \cite{Kumar2013} to be the same as that of the
isotropic spin system \cite{Chakrabarti1999}.  Moreover, it was claimed that the above
exponents remain the same as the length of the DNA changes from $N=16$ to $32$ base pairs
\cite{Mishra2013jcp}. 

The studies of Kumar et al. were restricted to small chain lengths because their simulations
were computationally expensive due to the excluded volume interactions which limits the
smallest time scale. However, using Monte Carlo simulations we could simulate the DNA of
lengths up to $N=512$ to study its behavior under the influence of a periodic force.  For this
purpose, we used a ($1+1$) dimensional model of dsDNA.  This model can be solved exactly in the
static force limit, and it has been studied in great detail via the generating function, and
the exact transfer matrix techniques~\cite{Marenduzzo2001,Marenduzzo2002,Kapri2004}. It has
been found that the results obtained from this model agrees qualitatively with the experimental
results, and other models used for studying DNA unzipping (see ~\cite{Kumar2010} and references
therein). By using finite size scaling on the DNA of lengths $N=128$, $256$, and $512$, we find
that the area of the hysteresis loop scales as $1/\omega$ at high frequencies whereas it scales
as $G^{\alpha} \omega^{\beta}$ with $\alpha = 1$ and $\beta = 5/4$ in the low-frequency regime.
These exponents are completely different from the values $\alpha = \beta = 1/2$ reported in
Ref.~\cite{Kumar2013,Mishra2013jcp}. 

The paper is organized as follows: In Sec.~\ref{sec:model}, we define our model and compare it
with the model studied by Kumar et al.~\cite{Kumar2013,Mishra2013,Mishra2013jcp}. The results
are discussed in Sec.~\ref{sec:results}. We summarize our results in Sec.~\ref{sec:summary}.

\section{Model} \label {sec:model}

\figone

The model used in this paper has been used previously in Ref.~\cite{Kapri2012} to study
the hysteresis in DNA unzipping by changing the pulling rate. In this model, the two strands of
a homo-polymer DNA are represented by two directed self-avoiding walks on a ($d=1+1$)-dimensional
square lattice.  The walks starting from the origin are restricted to go towards the positive
direction of the diagonal axis (z-direction) without crossing each other. The directional
nature of the walks takes care of self-avoidance and the correct base pairing of DNA, i.e., the
monomers that are complementary to each other are allowed to occupy the same lattice site. For
each such overlap there is a gain of energy $-\epsilon$ ($\epsilon >0$).  One end of the DNA is
anchored at the origin and a time-dependent periodic force
\begin{equation}
	\label{eq:1}
	g(t) = G \left| \sin \left( \omega t \right) \right|,
\end{equation}
with angular frequency $\omega$ and amplitude $G$ acts along the transverse direction ($x$
direction) at the free end. The strands of the DNA cannot cross each other, therefore, in the
negative cycle of the sine function the strands remain in the zipped state. By taking the
absolute value of the sine function in Eq.~(\ref{eq:1}), we have converted the negative cycles
to positive, thus reducing the time period by half. Hence the angular frequency of the external
force is $\omega = \pi \nu$ ($\nu$ is the linear frequency). Throughout the paper, by frequency
we mean the angular frequency.  The schematic diagram of the model is shown in
Fig.~\ref{fig:1}.

In the limit $\omega \to 0$, i.e., the static force limit, this model can be solved
exactly via a generating function and the exact transfer matrix techniques. It has been used
previously to obtain the phase diagrams of the DNA unzipping
\cite{Marenduzzo2001,Marenduzzo2002,Kapri2004}. For the static force case, the temperature
dependent phase boundary is given by
\begin{equation}
	g_c(T) = - \frac{T}{2} \ln \lambda (z_2),
	\label{eq:gc}
\end{equation}
where $\lambda(z) = (1- 2z -\sqrt{1-4z})/(2z)$ and $z_2 = \sqrt{1 -e^{-\beta \epsilon} } - 1 +
e^{-\beta \epsilon}$. The zero force melting takes place at a temperature
$T_m = \epsilon/\ln (4/3)$ (for details see Ref.~\cite{Kapri2012}). In this paper we will be
working at temperature $T=1$, and from Eq.~(\ref{eq:gc}) we get the critical force $g_c(1) =
0.6778\dots$. 

We perform Monte Carlo simulations of the model by using the METROPOLIS algorithm. The strands of
the DNA undergo Rouse dynamics that consists of local corner-flip or end-flip
moves~\cite{Doi1986} that do not violate mutual avoidance (the self-avoidance is taken care of by
the directional nature of the walks). The elementary move consists of selecting a random
monomer from a strand, which itself is chosen at random, and flipping it. If the move results
in overlapping of two complementary monomers, thus forming a base-pair between the strands, it
is always accepted as a move.  The opposite move, i.e. the unbinding of monomers, is chosen
with the Boltzmann probability $\eta = \exp(- \epsilon/ k_B T)$. If the chosen monomer is
unbound, whatever remains unbound after the move is performed is always accepted.  The time is
measured in units of Monte Carlo Steps (MCS).  One MCS consists of $2N$ flip attempts, i.e., on
an average, every monomer is given a chance to flip. Throughout the simulation, the detailed
balance is always satisfied. From any starting configuration, it is possible to reach any other
configuration by using the above moves.  Throughout this paper, without loss of generality, we
have chosen $\epsilon = 1$ and $k_B = 1$. 

At any given frequency $\omega$ and the force amplitude $G$, if the time $t$ is incremented by
unity, the external force $g(t)$ changes, according to Eq.~(\ref{eq:1}), from $0$ to a maximum
value $G$ and then decreases to $0$. Between each time increment, the system is relaxed by a
unit time (1MCS). Upon incrementing $t$ further, the above cycle gets repeated again and again.
Before taking any measurement, the simulation is run for $2000$ cycles so that the system can
reach the stationary state.  

At this point, it is worthwhile to compare our model with the model of Kumar et
al.~\cite{Kumar2013,Mishra2013,Mishra2013jcp}. In their model, a chain of length $N$, whose
first $N/2$ monomers are complementary to the rest half, is anchored (at origin) from one end,
and a periodic force is acting on the free end along $x-$direction. The monomers of the chain
are chosen in such a manner that the $i$th monomer from the anchored end can bind only wit theh
$(N-i)$th monomer of the chain, thus mimicking the base pair of the DNA. The system evolves in
the presence of an external periodic force, and the distance of the end monomer from the
origin, $x(t)$, is monitored as a function of time by using Langevin dynamics simulation. If
$x(t) < 5$ (for $N=32$) the DNA is taken to be in the zipped state, otherwise it is in the
unzipped state. Their model becomes similar to ours (see Fig.~\ref{fig:1}) if, instead of
first, the bead at the center of the chain is anchored, and a periodic force is applied on the
first and the last monomers in opposite directions. In both models, the forcing is such that
the force, averaged over a cycle, applied on the DNA is not equal to zero. Therefore, we expect
that for a given amplitude $G$, both models will have similar steady states in the larger
frequency limit. This is indeed the case. For lower values of $G$ (e.g., $G=0.4$ in
Ref.~\cite{Kumar2013} and $G=1$ in our case), the steady state is a zipped configuration, while
for higher values of $G$ (e.g., $G=1$ in Ref.~\cite{Kumar2013} and $G=3$ in our case), the DNA
is in an unzipped state. There are, however, few differences between the two models, but that
has more to do with the simulation technique. For example, in Langevin dynamics simulations,
friction needs to be introduced to equilibrate the system. In contrast, Monte Carlo dynamics is
dissipative by definition and brings the system to equilibrium.  Even though the Langevin
dynamics simulations of Kumar et al. are done in 3 dimensions, the motion of the end bead is
along the direction of an externally applied force. The quantity of interest is the hysteresis
traced out by the end monomer. For temperatures below the melting temperature of DNA, the
fluctuations of the end bead along the transverse directions are small and hence the area of
the hysteresis loop in the transverse directions is negligible. We have checked this for a self
avoiding polymer in three dimensions.  Therefore, our two-dimensional model captures the
essential physics of dynamic transitions, and we expect values for the exponents $\alpha$ and
$\beta$ similar to those obtained by Kumar et al.~\cite{Kumar2013,Mishra2013jcp}.  

\figtwo

We monitor the distance between the end monomers of the two strands as a function of time,
$x(t)$, for various force amplitudes $G$ and frequency $\omega$. The time
averaging of $x(t)$ over a complete period,
\begin{equation}
	Q = \frac {\omega}{\pi} \oint x(t) dt,
\end{equation}
may be used as a dynamical order parameter~\cite{Chakrabarti1999}. Since the force is
periodic in nature, we obtain the extension $x(g)$ as a function of force $g$ from the
time series $x(t)$, and we average it over $10^4$ cycles to obtain the average extension
$\langle x(g) \rangle$. If the force amplitude $G$ is not very small, and the frequency
$\omega$ of the periodic force is sufficiently high to avoid equilibration of the DNA, the
average extension, $\langle x(g) \rangle$, for the forward and the backward paths is not the same
and we see a hysteresis loop. The area of the hysteresis loop, $A_{loop}$, defined by
\begin{equation}
	A_{loop} = \oint  \langle x(g) \rangle dg,
	\label{eq:area}
\end{equation}
depends upon the frequency $\omega$ and the amplitude $G$ of the oscillating force and also
serves as another dynamical order parameter~\cite{Chakrabarti1999}. We bin the data generated
according to values of $g$ by using Eq.~(\ref{eq:1}). We first divide the interval $g \in [0,
G]$, for both the rise and fall of the cycle, into $1000$ uniform intervals, and we obtain the
value of $\langle x(g) \rangle$ at the end points of these intervals by interpolation using
cubic splines of the GNU Scientific Library~\cite{Galassi2009}.  The area of the loop
$A_{loop}$ is then evaluated numerically by using the trapezoidal rule on these intervals.

In this paper we report the behavior of $A_{loop}$ at high and low frequencies at various
force amplitudes $G$. The results for the dynamical order parameter $Q$ will be published
elsewhere~\cite{KapriUnpub}.

\section{Results and Discussions} \label{sec:results}

In  Fig.~\ref{fig:2}, we have shown the time variation of external force $g(t)$ and the scaled
extension $x(t)/N $ of different monomers for the DNA of length $N=128$ at various $G$ and
$\omega$ values for three consecutive cycles. This figure gives us important information that
can be used to estimate the frequency $\omega^{*}(G)$ at which the loop area $A_{loop}$ is
maximum. From Eq.~(\ref{eq:1}), one can see that for $T=1$, the number of time steps required
(say $t_z$) for a given frequency $\omega$ to increase $g(t)$ above the critical force $g_c(1)$
are approximately $\pi/4\omega$ and $\pi/12\omega$ for $G=1$ and $3$, respectively. Therefore,
for this time, the DNA remains in a zipped state.  The time required to unzip the DNA is given
by $t_{u} \sim N$. Since the magnitude of the force continues to increase much beyond $g_c$, the
DNA also keeps on stretching until it reaches a fully stretched configuration. This takes the
time $t_s \sim N$. Assuming that the DNA takes the same time ($t_u + t_s$) in reaching a zipped
configuration from the fully stretched unzipped state, the total time which sets the time scale
of the dynamics of the DNA is 
\begin{equation}
	t_{tot} = 2(t_z + t_u + t_s) =  \begin{cases}
		2 \left( 2N + {\pi}/{4\omega} \right) & \text{for} \ G = 1 \\
		2 \left( 2N + {\pi}/{12\omega} \right) &\text{for} \ G = 3, 
	\end{cases}
\end{equation}
for two different force amplitudes. If $t_{tot}$ matches with the time period of the
oscillating force (see Figs.~\ref{fig:2}(a) and \ref{fig:2}(b)), we get the maximum loop area. This
happens at the frequency
\begin{equation}
	\omega = \omega^{*}(G) = \begin{cases}
		{\pi}/{8 N} & \text{for} \ G = 1 \\
		{5 \pi}/{24 N} & \text{for} \ G = 3.
	\end{cases}
	\label{eq:w0}
\end{equation}
The values of $\omega^{*}(G)$ calculated from the above equation for various lengths of the DNA
are tabulated in Table~\ref{table:1}. 

\tableone

\figthree

In Fig.~\ref{fig:3}(a) we have plotted the area of the hysteresis loop, $A_{loop}$, as a
function of frequency $\omega$, for the DNA of length $N=128$ at force amplitudes $G=1$ and
$3$. The plot shows that the area of the hysteresis loop is a non monotonic function of the
frequency, and its behavior depends on the amplitude of the periodic force $G$. For the
equilibrium case, i.e., $\omega=0$, there is no hysteresis, resulting in a zero loop area. For
very low frequencies, the force changes very slowly, the DNA gets enough time to relax to this
change and it remains in equilibrium, so the loop area $A_{loop}$ is very small. Upon
increasing $\omega$, the change in the force in unit time increases, and there is some lag in
the response of the DNA to this change. This is depicted by an increase in the area of the
hysteresis loop.  The increase in $A_{loop}$ does not continue indefinitely with an increase in
$\omega$.  There is a frequency $\omega^{*}(G)$ ($\omega^{*}(G) \approx 2.8 \times 10^{-3}$ and
$4.2 \times 10^{-3}$, respectively, for $G=1.0$ and $3.0$) at which $A_{loop}$ is maximum and
it starts decreasing upon increasing $\omega$ above $\omega^{*}(G)$.  For $G=1$, $A_{loop}$
decreases monotonically as $\omega$ increases above $\omega^{*}(G)$, whereas it shows an
oscillatory component for the force amplitude $G=3$. The observed values of $\omega^{*}(G)$
obtained from the simulation are also tabulated in Table~\ref{table:1}. These values match
reasonably well with the frequencies calculated using Eq.~(\ref{eq:w0}).  

To understand the different behavior of $A_{loop}$, we have plotted the average extension
$\langle x(g) \rangle$, as a function of the applied force $g$ in Figs.~\ref{fig:3}(b)-(c) for
$G=1$ and Figs.~\ref{fig:3}(d)-(e) for $G=3$, for various frequencies marked in
Fig.~\ref{fig:3}(a) by capital letters ($A$ to $F$). For $G=1$, which lies slightly above the
critical force, $g_c$, needed to unzip the DNA ($g_c(1) = 0.6778\ldots$ from
Eq.~(\ref{eq:gc})), the majority of bonds of the DNA are in the zipped state (i.e., $\langle x
\rangle / N << 1$) when $g=0$.  (see Fig.~\ref{fig:3}(b)). When the force changes from $g=0$ to
$g=G=1$ very rapidly (point $A$ in Fig.~\ref{fig:3}(a) which corresponds to $\omega = 7.14
\times 10^{-2}$) then the fluctuating force can open only a few base pairs of the zipped DNA at
the open end and the area of the hysteresis loop is small. As $\omega$ decreases from
$\omega_{A} = 7.14 \times 10^{-2}$ (point $A$) to $\omega_{F} = 1.57 \times 10^{-2}$ (point
$F$), the DNA gets more time to relax, more and more base pairs become open, and the area of
the hysteresis loop increases. Figure~\ref{fig:3}(c), shows the hysteresis curve having a
maximum loop area at frequency $\omega^{*}(G=1) = 2.8\times 10^{-3}$.  Similar types of
hysteresis loops are also observed for the amplitude $G=0.65$, which lies below the phase
boundary. In this case, the majority of the bonds of the DNA are in the zipped state. However,
at any finite temperatures, a few bonds at the end become open due to thermal fluctuations. The
free ends are then dragged by the pulling force, resulting in a hysteresis loop with a small
area. 

The situation for the larger force amplitudes (see Fig.~\ref{fig:3}(a) for $G=3$) is however
different. The force $g = G$ lies far away from the phase boundary and at this force
value, the DNA is in the unzipped phase with a completely stretched conformation in the steady
state. At a very high frequency $\omega_{A} = 7.14 \times 10^{-2}$ (point $A$), the force
changes rapidly between $g=0$ and $g=G=3$ and DNA does not get enough time to respond to this
change. The separation between the end monomers remains constant, resulting in a small loop area.
Upon decreasing the frequency $\omega$, the DNA gets more time to relax. As a result the area of
the loop starts increasing.  However, it increases only up to frequency $\omega_{B} = 5.61
\times 10^{-2}$ (point $B$) and then decreases again till $\omega_{C} = 4.62 \times 10^{-2}$
(point $C$) and so on. This behavior continues up to $\omega^{*}(G=3) = 4.2 \times 10^{-3}$
for which we get the highest peak. Upon decreasing the frequency further the loop area decreases
and becomes zero in the limit $\omega \to 0$.  

The hysteresis loops (shown in Fig.~\ref{fig:3}(d) by filled diamonds, squares, and circles) at
frequencies $\omega_{A} = 7.14 \times 10^{-2}$, $\omega_{C} = 4.62 \times 10^{-2}$ and
$\omega_{E} = 2.24 \times 10^{-2}$ (points $A$, $C$ and $E$, respectively), where $A_{loop}$
has a minimum, have a peculiar shape. These loops have almost the same extension $\langle x(g)
\rangle$ at the minimum ($g = 0$) and the maximum ($g = G =3$) force values, and their shapes
are symmetrical about the line joining them. The loops at frequencies $\omega_{B}$ and
$\omega_{D}$ (point $B$ and $D$), where $A_{loop}$ has a maximum, are not symmetrical.  The
hysteresis curve having the maximum loop area at frequency $\omega^{*}(G=3) = 4.2\times
10^{-3}$ is also shown in Fig.~\ref{fig:3}(e).

The simple analysis used at the beginning of this section to calculate the frequency
$\omega^{*}(G)$ can be extended to estimate the frequency $\omega_{E}$, where the first minimum
arises for the force amplitude $G=3$ (see Fig.~\ref{fig:3}(a)). It can be seen in
Fig.~\ref{fig:2}(c) that for $G=3$, the 32nd monomer is in the bound state for lower values of
$g(t)$. This means that a fraction, $N/4$ of the length of the DNA from the anchored end is in
the zipped state. As the value of $g(t)$ increases, this bound segment of the DNA unzips. The
time required to unzip this fraction is $t_u \sim N/4$.  The kink that is generated as a
result of unzipping has to travel to the free end so that the DNA can take the stretched
configuration. It takes the time proportional to the length of the unbound segment of the DNA,
and $t_s \sim 3N/4$. Therefore, the total time for this case is $ t_{tot} = t_z + t_u + t_s
\approx  \pi/12\omega + N$. If this time is equal to the time period of the external force, the
DNA is out of phase with the external frequency and we get a minimum, giving the frequency
\begin{equation} 
	\omega = \omega_{E} = \frac{11 \pi}{12 N} \quad \text{for} \ G = 3,
	\label{eq:wE} 
\end{equation} 
at which $A_{loop}$ has the first minimum from the lower frequency side. The values of
$\omega_{E}$ are also tabulated in Table~\ref{table:1}. It matches excellently with the
observed frequency. The above analysis is limited by the estimation of the fraction of zipped
monomers of the DNA. This length decreases rapidly upon increasing the frequency, and its
estimation becomes more and more difficult. Taking again the fraction, $N/4$ of the length of
the DNA from the anchored end to be in the zipped state (see Fig.\ref{fig:2}(d)), we estimate
\begin{equation}
	\omega = \omega_D \approx \frac{5 \pi}{3 N} \quad \text{for} \ G=3,
	\label{eq:wD}
\end{equation} 
as the frequency of the second peak. This estimate has a deviation of around $20\%$ from the
frequencies observed from the simulation data, which may be due to the error in the estimation
of the length of the zipped segment of the DNA.

In Fig.~\ref{fig:4}, we have plotted the area of the hysteresis loop, $A_{loop}$, as a function
of $\omega$ for various force amplitudes $G$ ranging from $0.65$ to $3.0$.  The value $G=0.65$
lies just below the phase boundary $g_c(T)$ (given by Eq.~(\ref{eq:gc})) for the static case
($\omega = 0$), and all other values lie above it.  For smaller values of $G$, the area keeps
on decreasing for frequencies above $\omega^{*}(G)$ and eventually becomes zero at very high
frequencies.  For $G=0.65$, which lies below the phase boundary, the loop area is of the order
$1$. We have scaled it by a factor of $10$ to make it visible in the plot.  Therefore, in the
limit $N \to \infty$, the loop area per unit length, $A_{loop}/N \to 0$ for values of $G$ that
lies below the phase boundary. For higher values of $G$, however, we find that the area of the
loop has an oscillatory component. It shows few more peaks of smaller heights before going to
zero at very high frequencies. One such peak is shown in the inset of Fig.~\ref{fig:4} between
the frequency range $2.0 \times 10^{-2}$ and $4.5 \times 10^{-2}$.  One can clearly see that,
in this frequency range, for $G=1.0$ and $1.25$, the loop area decreases but for $G=1.5$ and
higher it first increases, reaches a local maximum, and then decreases. For a given amplitude
$G$, we found that the number of peaks depends on the length $N$ of the DNA. Similar oscillatory
behavior is also observed in the other order parameter $Q$~\cite{KapriUnpub}.

\figfour

\figfive

In Figs.~\ref{fig:5}(a) and \ref{fig:5}(c), we have plotted the area of the hysteresis loop,
$A_{loop}$, as a function of frequency $\omega$ for the DNA of lengths $N=32$, 64, 128, 256,
and 512 at force amplitudes $G = 1$ and $3$, respectively, in a log-log scale. In the high
frequency range the loop area decreases linearly for $G=1$ as $\omega$ increases, whereas it
shows oscillatory behavior for $G=3$. The number of peaks increases as the length of the DNA
increases.  We explored up to the frequency $\omega = 3.14 \times 10^{-1}$ and found that for
the DNA of length $N=32$ only one secondary peak exists, whereas for $N=512$, there are 15
peaks. Such oscillatory behavior were not not seen in earlier
studies~\cite{Kumar2013,Mishra2013jcp} because they were done on short DNA (the maximum length
used was only $N=32$ base pairs), with frequencies much lower than reported in this paper. In
the lower frequency range, it is clearly visible that the slope of $A_{loop}$ changes as the
length of the DNA increases. The above behavior shows the presence of strong finite size
effects, and the exponents obtained by finite-size scaling with lengths up to
$N=32$~\cite{Mishra2013jcp} needs to be estimated again using longer chain lengths.   

From Figs.~\ref{fig:5}(a) and \ref{fig:5}(c), it is clear that $\omega^{*}(G)$, the frequency
at which the loop area is maximum, decreases as the length of the DNA is increased. In the
thermodynamic limit $N \to \infty$, from Eq.~(\ref{eq:w0}), we get $\omega^{*}(G) \to 0$. This
suggests the scaling form for the loop area $A_{loop}$,
\begin{equation} 
	A_{loop} = N^{\delta} \mathcal{G} \left( \omega N^{z} \right),
	\label{eq:Aloop} 
\end{equation} 
where $\delta$ and $z$ are critical exponents. The exponent $z$ is the dynamic exponent as time
$t \sim N^{z}$. We obtain a nice data collapse for $\delta =1.06 \pm 0.05$ and $z=1.05 \pm
0.03$ for $G=1$, and $\delta=1.02 \pm 0.02$ and $z=1.01 \pm 0.01$ for $G=3$. The data collapse
for $G=1$ and $3$ are shown in Figs.~\ref{fig:5}(b) and \ref{fig:5}(d), respectively.  The
scaled curve of Fig.~\ref{fig:5}(d) clearly shows an oscillatory component in the loop area for
higher force amplitudes. In the following we explain the reason for getting $z=1$. For an ideal
Rouse chain of length $N$, the longest relaxation time (Rouse time) $\tau_{R} \sim N^2$.  On
the time scale $t > \tau_R$, the motion of the chain is diffusive, i.e., the mean-square
displacement is linear in time, giving $z=2$. For a constant pulling force above the phase
boundary, the time $t$ required to unzip the DNA of length $N$ from a nonequilibrium zipped
state to an unzipped state at equilibrium is found to be $\sim
N^2$~\cite{Sebastian2000,Marenduzzo2002}.  In recent years, different dynamical exponents for
zipping time have been found in various DNA zipping
simulations~\cite{Ferrantini2011,Dasanna2012}. For example, an anomalous exponent of $z=1.37$
has been found in the simulations of zipping dynamics of two flexible polymers anchored at one
end by Ferrantini and Carlon~\cite{Ferrantini2011}.  In another study, Dasanna et
al.~\cite{Dasanna2012} have simulated a semiflexible model of DNA that explicitly includes the
bending rigidities of dsDNA segments, and they found the value $z=1.4$ for the zipping time.   
For the present problem, however, the frequency of the external force is such that its time
period is much smaller than $\tau_R$.  The chain never relaxes completely, and its motion is
subdiffusive, i.e., the mean square displacement increases as the square root of
time~\cite{Rubinstein2003}.  Our model also allows fluctuations in the length of the DNA. For
lower values of force $g$, the DNA is in the zipped state, where it takes a zig-zag
configuration. However, for higher values of $g$, the DNA is in the unzipped state with a fully
stretched configuration. The average length of the DNA in the unzipped state is more than its
length in the zipped state. Due to these length fluctuations, we also have longitudinal modes
of the Rouse chain.  For $t < \tau_{R}$, the mean squared contour length of the chain increases
as the square root of time ~\cite{Rubinstein2003} and therefore $z=1$. Due to the geometry of
the square lattice, the change in length of the DNA by flipping a monomer (diagonal along
$z-$axis) is exactly equal to the change in the separation of the end monomers (diagonal along
$x-$axis).  Hence the end-separation correlation function $\langle x(t) x(0) \rangle$ is
exactly equal to the length correlation function and should scale as $t/N$. This is indeed
found in the simulation giving $z=1$~\cite{KapriUnpub}.  The exponents $\delta=1$ and $z=1$ are
similar to the exponents obtained in Ref.~\cite{Mishra2013jcp} using DNA of shorter lengths.

A Rouse chain of length $N$ has natural frequencies at $(2p-1)\pi/2N$, where
$p=1,2,\ldots$ are integers. When this frequency matches with the frequency $\omega$ of
the externally applied periodic force, we get a resonance. From Fig.~\ref{fig:5}(d), one
can see that for $G=3$, the location of maxima (minima) is situated when the scaled
frequency $\omega N$ is an odd (even) integral multiple of $\pi/2$. Therefore, the length
dependent frequency $\omega_{p}$ of these maxima (minima) is given by
\begin{equation}
	\omega_{p} = \begin{cases}
		{(2p -1) \pi}/{2N} & \text{(maxima)} \cr
		{p \pi}/{N} & \text{(minima)},
	\end{cases}
	\label{eq:modes}
\end{equation}
where $p=1,2,\ldots$ are integers. From the above expression, the first peak, which
corresponds to the maximum loop area, has a frequency $2.5$ times higher than that
predicted by Eq.~(\ref{eq:w0}).  However, the frequencies of the higher peaks and valleys
that are estimated by Eq.~(\ref{eq:modes}) are quite close to those observed in the
simulation. This is because these modes, as opposed to the first mode, get completely
relaxed within the time period of the applied force. The values of the second mode
$\omega_{2}$ for various $N$ are also tabulated in Table \ref{table:1}. These matches
extremely well with the observed frequencies.

\figsix

In Fig.~\ref{fig:6}, we have plotted the area of the hysteresis loop, $A_{loop}$, as a function
of $G^{\alpha} \omega^{\beta}$ in the low frequency range for the DNA of lengths $N=128$, 256,
and 512 at various force amplitudes $G$. A good data collapse is obtained for the values
$\alpha = 1.0 \pm 0.05$ and $\beta=1.25 \pm 0.05$.  The values of $\alpha$ and $\beta$ differ
considerably with previously obtained values $\alpha = \beta =
1/2$~\cite{Kumar2013,Mishra2013jcp}. We believe that the lower values of exponents are due to
the shorter chain lengths used in their simulations.

The scaling function $\mathcal{G}(\omega)$ can be obtained by observing that at low frequencies
($\omega \to 0$), for large $N$, the $A_{loop}$ scales as $G^{\alpha}\omega^{\beta}$, while at
very high frequencies ($\omega \to \infty$), from Eq.~(\ref{eq:Aloop}), we see $A_{loop} \sim 1
/ \omega$. For smaller $G$ values the steady state is a zipped configuration, and the area of
the loop decreases monotonically for $\omega$ above $\omega^{*}(G)$ (e.g., $G=1$). For such
cases, the scaling function $\mathcal{G}(\omega)$ that satisfies the above requirements has the
form
\begin{equation}
	\mathcal{G}(\omega) = \frac{A G^{\alpha} \omega^{\beta}}{\omega^{1+\beta} + B^2},
	\label{eq:sfun}
\end{equation}
with $A$ and $B$ as fitting parameters. The scaling function for $G=1$, with parameters
$A = 0.36$ and $B=0.30$, obtained by data fitting, is plotted in Fig.~\ref{fig:5}(b).
For moderate force amplitudes (say $G=1.25$), we found that the above scaling function
is still valid with parameters $A=0.78$ and $B= 0.39$.
 
For higher force amplitudes, the steady state is a completely stretched unzipped state, and the
loop area has an oscillatory component above $\omega^{*}(G)$. For such cases, the above form is
not suitable as the scaling function.

\section{Conclusions} \label{sec:summary}

In this paper we have studied the hysteresis in unzipping of a dsDNA by a periodic force with
frequency $\omega$ and amplitude $G$ for chains up to lengths $N=512$ by using Monte Carlo
simulations.  The behavior of the loop area depends on the force amplitudes.  We find that for
lower $G$ values, the steady state of the DNA is a zipped configuration. The area of the loop
shows only one peak at $\omega^{*}(G)$, and for frequencies above $\omega^{*}(G)$, it decreases
monotonically. However, for higher force amplitudes, the steady state is an unzipped state and
the area of the loop shows multiple peaks. We gave a simple analysis that could estimate
$\omega^{*}(G)$, the frequency at which the maximum loop area is observed, and the frequencies
of other peaks that appear for higher force amplitudes.  We also explored the behavior of the
hysteresis loop area for a wide range of frequencies for both lower and higher values of force
amplitudes $G$ using the finite-size scaling. We found that the loop area scales as $1/\omega$
in the high-frequency range, whereas it scales as $G^{\alpha}\omega^{\beta}$ with exponents
$\alpha=1$ and $\beta=5/4$ in the low-frequency regime. These exponents are found to be
different from the values obtained by Kumar et al.~\cite{Kumar2013,Mishra2013jcp}. We believe
that the different values of exponents $\alpha$ and $\beta$ are due to the shorter chain
lengths used in their studies. It would be interesting to study longer chain lengths using
Langevin dynamics simulations to confirm the above results.  The other interesting direction
would be to include the bending rigidity of dsDNA that has been ignored in this paper and study
its influence on the dynamic exponents as a function of chain lengths. In fact, for chains of
the order of persistence length of the DNA, the bending rigidity could play an important role
and the dynamic exponents may be different than reported in this paper.  However, in the
thermodynamic limit, the bending rigidity would not be relevant and we believe that the
exponents of the flexible chain will be recovered. This, however, will require further
investigations.

\section*{Acknowledgements}

I thank A. Chaudhuri for discussions and D. Dhar and S. M.  Bhattacharjee for their valuable
comments and suggestions on the manuscript. I acknowledge the HPC facility at IISERM for
generous computational time. This work is supported by DST (India) Grant No.
SR/FTP/PS-094/2010.

\end{document}